# Quantum Nash Equilibria and Quantum Computing

**Philip Vos Fellman**
Southern New Hampshire University
Manchester, New Hampshire
Shirogitsune99@yahoo.com

**Jonathan Vos Post**
Computer Futures, Inc.,
Altadena, California
Jvospost2@yahoo.com

## 1.1 Introduction: The Complexity of Nash Equilibrium

In 2004, At the Fifth International Conference on Complex Systems, we drew attention to some remarkable findings by researchers at the Santa Fe Institute (Sato, Farmer and Akiyama, 2001) about hitherto unsuspected complexity in the Nash Equilibrium. As we progressed from these findings about heteroclinic Hamiltonians and chaotic transients hidden within the learning patterns of the simple rock-paper-scissors game to some related findings on the theory of quantum computing, one of the arguments we put forward was just as in the late 1990's a number of new Nash equilibria were discovered in simple bi-matrix games (Shubik and Quint, 1996; Von Stengel, 1997, 2000; and McLennan and Park, 1999) we would begin to see new Nash equilibria discovered as the result of quantum computation.

While actual quantum computers remain rather primitive (Toibman, 2004), and the theory of quantum computation seems to be advancing perhaps a bit more slowly than originally expected, there have, nonetheless, been a number of advances in computation and some more radical advances in an allied field, quantum game theory (Huberman and Hogg, 2004) which are quite significant. In the course of this paper we will review a few of these discoveries and illustrate some of the characteristics of these new "Quantum Nash Equilibria".

Much of the recent development in quantum computing might be likened to the early efforts at developing electronic music. As Toibman (2004) points out, the current generation of quantum computers are so small that they cannot yet store enough information to maintain the word "Hello" in their registers. This is reminiscent of a story that composer Alden Ashforth (Byzantium, Sailing to Byzantium: Two Journeys after Yeats) tells of Milton Babbitt demonstrating the electronic music laboratory at Princeton in the 1960's. After a lecture on acoustics and the demonstration of several million dollars of digital and analog equipment, Babbitt's visitor asked repeatedly (and apparently irritatingly) "I want to hear an oboe". Despite the complex array of equipment, the production of a single 'lifelike' note mimicking a musical instrument was not possible at the time. Forty years later, convincing digital simulations of entire orchestras are routine work and quite complex musical synthesis is possible even with inexpensive home devices. A similar pattern of development would not be unreasonable to expect for quantum computation and quantum computational devices.

## 1.2 Meyer's Quantum Strategies: Picard Was Right

In 1999, David Meyer, already well known for his work on quantum computation and modeling (Meyer, 1997) published an article in Physical Review Letters entitled "Quantum Strategies" which has since become something of a classic in the field (Meyer, 1999). In this paper, in a well known, if fictional setting, Meyer analyzed the results of a peculiar game of coin toss played between Captain Jean Luc Picard of the Starship Enterprise and his nemesis "Q". In this game, which Meyer explains as a two-person, zero-sum (noncooperative) strategic game, the payoffs to Picard and "Q" (P and Q hereafter) are represented by the following matrix showing the possible outcomes after an initial state (heads or tails) and two flips of the coin (Meyer, 1999):

|   | $NN$ | $NF$ | $FN$ | $FF$ |
|---|------|------|------|------|
| $N$ | $-1$ | $1$ | $1$ | $-1$ |
| $F$ | $1$ | $-1$ | $-1$ | $1$ |

…The rows and columns are labeled by P's and Q's pure strategies respectively; F denotes a flipover and N denotes no flipover, and the numbers in the matrix are P's payoffs, 1 indicating a win and -1 indicating a loss (p. 1052)

Meyer notes that this game has no deterministic solution and that there is no deterministic Nash equilibrium. However, he also notes (following von Neumann) that since this is a two-person, zero sum game with a finite number of strategies there must be a probabilistic Nash equilibrium which consists of Picard randomly flipping the penny over half of the time and Q randomly alternating between his four possible strategies. The game unfolds in a series of ten moves all of which are won by Q. Picard suspects Q of cheating. Meyer's analysis proceeds to examine whether this is or is not the case.

Meyer's first step is go back and analyze the sequence of moves in its extensive form, an entirely orthodox approach. However, rather than illustrating the sequence with a binary tree, he uses the directed graph shown below, with the three vertices labeled H or T according to the state of the penny, with diagonal arrows representing a flipover and vertical arrows representing no flipover (p. 1053):

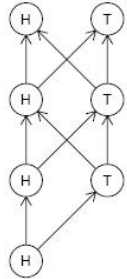

PQ Penny Flip in extensive form

He then converts the directed graph to a vector space V, with basis {H,T} representing player strategies by a series of 2 x 2 matrices (p.1053):

$$F := \begin{matrix} & H & T \\ H & \\ T & \end{matrix}\begin{pmatrix} 0 & 1 \\ 1 & 0 \end{pmatrix} \quad \text{and} \quad N := \begin{matrix} & H & T \\ H & \\ T & \end{matrix}\begin{pmatrix} 1 & 0 \\ 0 & 1 \end{pmatrix}$$

Meyer explains that in this notation, "A sequence of mixed actions puts the state of the penny into a convex linear combination $a$H + (1 - $a$)T, $0 \le a \le 1$, which means that if the box is opened, the penny will be head up with probability $a$." (p. 1053) It is at this point in the analysis that Meyer undertakes a bit of logical prestidigitation, arguing that "$Q$, however, is eponymously using a *quantum* strategy, namely a sequence of unitary, rather than stochastic matrices to act on the penny" (p. 1053) There follows then, an analysis, using standard Dirac notation of the quantum vector space and the series of unitary transformations on the density space which have the effect of taking Picard's moves (now defined not as a stochastic matrix on a probabilistic state, but rather as a convex linear combination of unitary, deterministic transformations on density matrices by conjugation) and transforming them by conjugation (Q's moves) .This puts the penny into a simultaneous eigenvalue 1 eigenstate of both F and N (invariant under any mixed strategy), or in other words, it causes the penny to "finish" heads up no matter what ("All of the pairs ([$p$F + (1 − $p$)N], | $U$ (1/√2,1/√2), $U$ (1/√2,1/√2)]) are (mixed, quantum) equilibria for *PQ* penny flipover, with value -1 to P; this is why he loses every game." (p. 1054).

## 1.3 Hadamards Unitary Matrices and Superpositioning

Why we refer to this series of actions as logical prestidigitation is because playing the quantum strategy in the PQ coin flipping game requires a *quantum* penny (i.e., a macroscopic object with quantum behavioral properties)[1]. The reader may not realize this upon first looking at statements of the problem.

---

[1]  It is this distinction which allows the penny to exhibit quantum mechanical properties and, in the case, allows the application of quantum game strategies. A lot of counter-intuitive or weird behavior comes out of this stipulation because objects at the quantum mechanical level don't behave like macroscopic objects. Feynman notes this when he explains how difficult it is to develop intuitions about the quantum-mechanical world and explains that he himself does not have intuitions per se of the quantum mechanical world (Feynman Lectures on Physics, Quantum Mechanics, Chapter 10). Toibman explains this in the PQ coin toss game (although he eschews the Star Trek example) in terms of twice Hadamarding the probability density matrix, noting that "It is interesting that before the second Hadamard is applied to the qubit, had it been observed, there would have been a 50% chance that we would have found tails; yet after the Hadamard operation is done again, the 50% chance disappears and becomes 0%. The net effect of the

Picard certainly didn't realize it, or he wouldn't have agreed to play once, much less ten times (which he does in order to capture a probabilistic c outcome in what is, in fact, not a probabilistic game). One of the reasons there are new Nash equilibria in the PQ penny flipping game is because it is a different penny, which can be in a simultaneous superpositioning of heads and tails. The quantum Nash equilibria in this game can also be explained by application of Hadamard to each of the two flips (Toibman, 2004).

The *PQ* coin flip does not, however, explore the properties of quantum entanglement. Landsberg (2004) credits the first complete quantum game to Eisert, Wilkens and Lewenstein (1999), whose game provides for a single, entangled state space for a pair of coins. Here, each player is given a separate quantum coin which can then either be flipped or not flipped. The coins start in the maximum entangled state:

$$\mathbf{H} \otimes \mathbf{H} + \mathbf{T} \otimes \mathbf{T}$$

Which in a two point strategy space allows four possible states (Landsberg, 2004):[2]

$$\mathbf{NN} = \mathbf{H} \otimes \mathbf{H} + \mathbf{T} \otimes \mathbf{T}$$

$$\mathbf{NF} = \mathbf{H} \otimes \mathbf{T} + \mathbf{T} \otimes \mathbf{H}$$

$$\mathbf{FN} = \mathbf{H} \otimes \mathbf{T} - \mathbf{T} \otimes \mathbf{H}$$

$$\mathbf{FF} = \mathbf{H} \otimes \mathbf{H} - \mathbf{T} \otimes \mathbf{T}$$

The strategies for this game can then be described in terms of strategy spaces which can be mapped into a series of quaternions. The Nash equilibria which occur in both the finite and (typically) non-finite strategy sets of these games can then be mapped into Hilbert spaces where, indeed new Nash equilibria do emerge as shown in the diagram from Cheon and Tsutsui below (Cheon and Tsutsui, 2006). As in the case of the quantum Nash equilibria for the *PQ* game, unique quantum Nash equilibria are a result of the probability densities arising from the action of self-adjoint quantum operators on the vector matrices which represent the strategic decision spaces of their respective games.

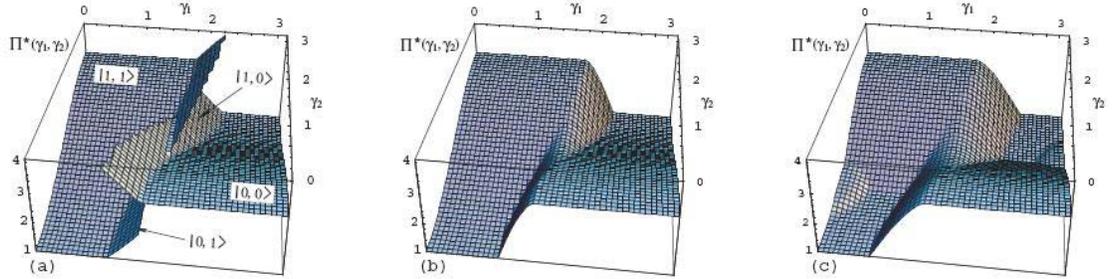

FIG. 1: The quantum Nash equilibrium payoff $\Pi_A^\star(\gamma) = \Pi_A(\alpha^\star, \beta^\star; \gamma)$ (or $\Pi_A(P_A^\star, P_B^\star; \gamma)$ for mixed quantum strategies) as a function of $\gamma$. Only the region $\gamma_1, \gamma_2 \in [0, \pi]$ is shown since $\Pi_A^\star(\gamma)$ has the reflection invariance $\Pi_A^\star(2\pi - \gamma_1, \gamma_2) = \Pi_A^\star(\gamma_1, 2\pi - \gamma_2) = \Pi_A^\star(\gamma_1, \gamma_2)$. The extra invariance $\Pi_A^\star(\pi - \gamma_1, \pi - \gamma_2) = \Pi_A^\star(\gamma_1, \gamma_2)$ is also visible. (a) Edge state Nash equilibria for $A_{00} = 3$, $A_{01} = 0$, $A_{10} = 5$ and $A_{11} = 1$. The value $(\gamma_1, \gamma_2) = (0.9272, 0)$ gives the maximum payoff $\Pi_A^\star = 4$ for one of the players. (b) Symmetric mixed quantum Nash equilibrium with the same parameters as (a). The maximum payoff $\Pi_A^\star = 3$ is obtained at $\gamma_2^\star = 0$ and $1.3694 \leq \gamma_1^\star < \pi$. (c) Mixed Nash equilibrium for $A_{00} = 3$, $A_{01} = 0$, $A_{10} = 5$ and $A_{11} = 0.2$. The two bumps near the left and right ends are due the pure symmetric Nash equilibria (20).

---

Hadamard is the same as adding $\frac{1}{\sqrt{2}}\binom{0}{-1}$. It contributes a − ½ probability to the coin coming up ⎪ tails). This is a basic example of quantum interference, where probabilities cancel or complement each other. *This is one of the weirdest aspects of quantum mechanics.*" (p. 8, Italics added)

[2] We have slightly altered Landsberg and Eisert, Wilkens and Lewenstein's notation to reflect that of Meyer's original game for purposes of clarity.

## 2.1 Realizable Quantum Nash Equilibria

Perhaps the most interesting area for the study of quantum Nash equilibria is coordination games. Drawing on the quantum properties of entangled systems quantum coordination games generate a number of novel Nash equilibria. Iqbal and Weigert (2004) have produced a detailed study of the properties of quantum correlation games, mapping both invertible and discontinuous g-functions and Non-invertible and discontinuous g-functions (as well as simpler mappings) arising purely from the quantum coordination game and not reproducible from the classical games.

Of possibly more practical interest is Huberman and Hogg's study (2004) of coordination games which employs a variant of non-locality familiar from the EPR paradox and Bell's theorem (also treated in detail by Iqbal and Weigert) to allow players to coordinate their behavior across classical barriers of time and space (see schematic below). Once again, the mathematics of entangled coordination are similar to those of the *PQ* quantum coin toss game and use the same kind of matrix which is fully elaborated in the expression of quantum equilibria in Hilbert space.

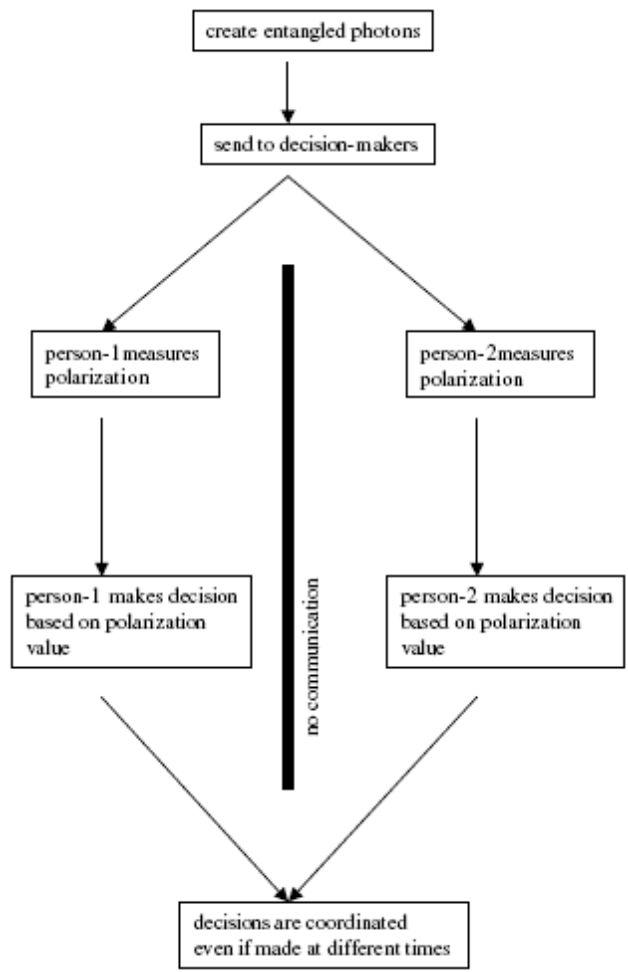

(Above) Huberman and Hogg's entanglement mechanism. In a manner similar to the experimental devices used to test Bell's theorem, two entangled quanta are sent to different players (who may receive and measure them at different times) who then use their measurements to coordinate game behavior. (Huberman and Hogg, 2004).

### 2.1.1 Quantum Entanglement and Coordination Games

In a more readily understandable practical sense, the coordination allowed by quantum entanglement creates the possibility of significantly better payoffs than classical equilibria. A quantum coordinated version of rock-paper-scissors, for example, where two players coordinate against a third produces a payoff asymptotic to 1/3 rather than 1/9. Moreover, this effect is not achievable through any classical mechanism since such a mechanism would involve the kind of prearrangement which would then be detectable through heuristics such as pattern matching or event history (Egnor, 2001). This kind of quantum Nash equilibrium is both realizable through existing computational mechanisms and offers significant promise for applications to cryptography as well as to strategy. As Huberman and Hogg note:

> As to the implementation of these mechanisms, these quantum solutions of coordination problems are not just a theoretical construct, as they can be implemented over relatively large distances. In particular, parametric-down conversion techniques can produce twin photons which are perfectly quantum correlated in time, space and often in polarization.) These photons can then be physically separated by many kilometers so that each participant gets one of the entangled particles. If the lifetime of the entangled state is long, each participant can then receive an entangled photon and perform a polarization measurement later, thus not having to communicate with each other during the whole procedure. On the other hand, if the lifetime of the entangled state is shorter than the period of the game, photons can be regenerated periodically, thereby requiring a transmission channel from the source to the participants (but not between the participants). In this case the advantage lies not in avoiding the possibility of blocked communication by an adversary, but in avoiding the detection of a coordinated solution and the direct communication among the participants (which is relevant when the participants wish to remain anonymous to the adversary). This makes for a feasible quantum solution to coordination problems that can be implemented with current technology, in contrast with most schemes for arbitrary quantum computation. (p. 429)

### 3.1 The Minority Game and Quantum Decoherence

Two other areas which we have previously discussed are the Minority Game, developed at the Santa Fe Institute by Challet and Zhang and the problem of decoherence in quantum computing. J. Doyne Farmer (Farmer, 1999) uses the Minority Game to explain learning trajectories in complex, non-equilibrium strategy spaces as well as to lay the foundation for the examination of complexity in learning the Nash equilibrium in the rock-paper-scissors game (Sato, Akiyama and Farmer, 2001). Adrian Flitney (Flitney and Abbot, 2005; Flitney and Hollenberg, 2005), who has done extensive work in quantum game theory combines both of these areas in a recent paper examining the effects of quantum decoherence on superior new quantum Nash equilibria.

### 3.1.1 The Minority Game

As Damien Challet explains:[3]

> A minority game is a repeated game where **N** (odd) players have to choose one out of two alternatives (say A and B) at each time step. Those who happen to be in the **minority** win. Although being rather simple at first glance this game is subtle in the sense that if all players analyze the situation in the same way, they all will choose the same alternative and will lose. Therefore, players have to be heterogeneous. Moreover, there is a frustration since not all the players can win at the same time: this is an essential mechanism for modeling competition. Note that this is an abstraction of the famous El-

---

[3] From "The Minority Game's Homepage", Damien Challet, http://www.unifr.ch/econophysics/minority/

Farol's bar problem…The Minority Game is simply a minority game with artificial agents with partial information and bounded rationality. They base their decision only on the knowledge of the **M** (M for memory) last winning alternatives, called **histories** ; there are $2^M$ histories. Take all the histories and fix a choice (A or B) for each of them : you get a **strategy**, which is like a theory of the world. Each strategy has an intrinsic value, called **virtual value**, which is the total number of times the strategy has predicted the right alternative. At the beginning of the game, every player gets a limited set of **S** strategies ; he uses them **inductively**, that is he uses the strategy with the highest virtual value (ties are broken by coin tossing). It must be emphasized that a player does not know anything about the others : all his informations come from the virtual values of the strategies.

### 3.1.2 The Minority Game as Information Processing

Both Flitney and Farmer[4] treat the minority game as an information processing system. Flitney and Hollenberg (2005)) explain:

A game can be considered an information processing system, where the players' strategies are the input and the payoffs are the output. With the advent of quantum computing and the increasing interest in quantum information it is natural to consider the combination of quantum mechanics and game theory. Papers by Meyer and Eisert et al. paved the way for the creation of the new field of quantum game theory. Classical probabilities are replaced by quantum amplitudes and players can utilize superposition, entanglement and interference. In quantum game theory, new ideas arise in two-player and multiplayer settings. In the protocol of Eisert et al, in two player quantum games there is no NE when both players have access to the full set of unitary strategies. Nash equilibria exist amongst mixed quantum strategies or when the strategy set is restricted in some way. Strategies are referred to as pure when the actions of the player at any stage is deterministic and mixed when a randomizing device is used to select among actions. That is, a mixed strategy is a convex linear combination of pure strategies. In multiplayer quantum games new NE amongst unitary strategies can arise. These new equilibria have no classical analogues.

### 3.2 Flitney and Abott's Quantum Minority Game

Flitney and Abbot then proceed through a brief literature review, explaining the standard protocol for quantizing games, by noting that "If an agent has a choice between two strategies, the selection can be encoded in the classical case by a bit." And that "to translate this into the quantum realm the bit is altered to a qubit, with the computational basis states $|0\rangle$ and $|1\rangle$ representing the original classical strategies." (p. 3) They then proceed to lay out the quantum minority game, essentially following the methodology used by Eisert, Wilkens and Lewenstein for the quantum prisoner's dilemma, specifying that (p.3):

The initial game state consists of one qubit for each player, prepared in an entangled GHZ state by an entangling operator $\hat{J}$ acting on $|00…0\rangle$. Pure quantum strategies are local unitary operators acting on a player's qubit. After all players have executed their moves the game state undergoes a positive operator valued measurement and the payoffs are determined from the classical payoff matrix. In the Eisert protocol this is achieved by applying $\hat{J}^\dagger$ to the game state and then making a measurement in the computational basis state. That is, the state prior to the measurement in the N-player case can be computed by:

---

$$\begin{aligned}
|\psi_0\rangle &= |00\ldots 0\rangle \\
|\psi_1\rangle &= \hat{J}|\psi_0\rangle \\
|\psi_2\rangle &= (\hat{M}_1 \otimes \hat{M}_2 \otimes \ldots \otimes \hat{M}_N)|\psi_1\rangle \\
|\psi_f\rangle &= \hat{J}^\dagger|\psi_2\rangle,
\end{aligned}$$

Where $|\psi_0\rangle$ is the initial state of the $N$ qubits, and $M_k$, $k = 1\ldots,N$ is a unitary operator representing the move of player $k$. The classical and pure strategies are represented by the identity and the flip operator. The entangling operator $\hat{J}$ continues with any direct product of classical moves, so the classical game is simply reproduced if all players select a classical move.

### 3.2.1 Decoherence

Flitney and Hollenberg explain the choice of density matrix notation for decoherence, and the phenomena which they are modeling (i.e., dephasing, which randomizes the relative phase between $|0\rangle$ and $|1\rangle$ states and dissipation which modifies the population of the states among other forms) explaining dephasing in terms of exponential decay over time of the off-diagonal elements of the density matrix and dissipation by way of amplitude damping. Decoherence is then represented following Eisert, Wilkens and Lewenstein (see Appendix III).

### 3.2.2. Quantum vs. Classical Minority Games

Flitney and Hollenberg begin their analysis of the quantum Minority game by noting that in the classical game, the equilibrium is trivial:

> In the classical Minority game the equilibrium is trivial: a maximum expected payoff is achieved if all players base their decision on the toss of a fair coin. The interest lies in studying the fluctuations that arise when agents use knowledge of past behaviour to predict a successful option for the next play. In the quantum game, as we shall see, a more efficient equilibrium can arise when the number of players is even. This paper only considers the situation where players do not make use of their knowledge of past behaviour. The classical pure strategies are then "always choose 0" or "always choose 1."

### 3.2.3 The Four Player Game

They then demonstrate Benjamin and Hayden's (2001) four player optimal quantum strategy result for the four player Minority game which results in a quantum Nash equilibrium with an expected payoff of ¼ to each player, which is the maximum possible for a symmetric strategy profile, and twice what can be achieved in the classical game (where players cannot do better than selecting 0 or 1 at random).

### 3.3 The Generalized Quantum Minority Game

Flitney and Hollenberg next generalize this result into a function δ, which yields a calculation of optimal strategy for generalized odd and even games with the interesting result that the:

> The NE that arises from selecting $\delta = \pi/(4N)$ and $\eta = 0$ may serve as a focal point for the players and be selected in preference to the other equilibria. However, if the players select $\hat{S}_{NE}$ corresponding to different values of n the result may not be a NE. For example, in the four player MG, if the players select $\eta_A$; $\eta_B$; $\eta_C$, and $\eta_D$ respectively, the resulting payoff depends on ($n_A + n_B + n_C + n_D$). If the value is zero, all players receive the quantum NE payoff of ¼, if it is one or three, the expected payoff is reduced to the classical NE value of 1/8 , while if it is two, the expected payoff vanishes. As a result, if all the players choose a random value of η the expected payoff is the same as that for the classical game (1

8 ) where all the players selected 0 or 1 with equal probability. Analogous results hold for the quantum MG with larger numbers of players.

When N is odd the situation is changed. The Pareto optimal situation would be for (N - 1 ) = 2 players to select one alternative and the remainder to select the other. In this way the number of players that receive a reward is maximized. In the entangled quantum game there is no way to achieve this with a symmetric strategy profile. Indeed, all quantum strategies reduce to classical ones and the players can achieve no improvement in their expected payoffs.

The NE payoff for the N even quantum game is precisely that of the N - 1 player classical game where each player selects 0 or 1 with equal probability. The effect of the entanglement and the appropriate choice of strategy is to eliminate some of the least desired final states, those with equal numbers of zeros and ones. The difference in behaviour between odd and even N arises since, although in both cases the players can arrange for the final state to consist of a superposition with only even (or only odd) numbers of zeros, only in the case when N is even is this an advantage to the players. Figure 5 shows the maximum expected payoffs for the quantum and classical MG for even N. (p. 8)

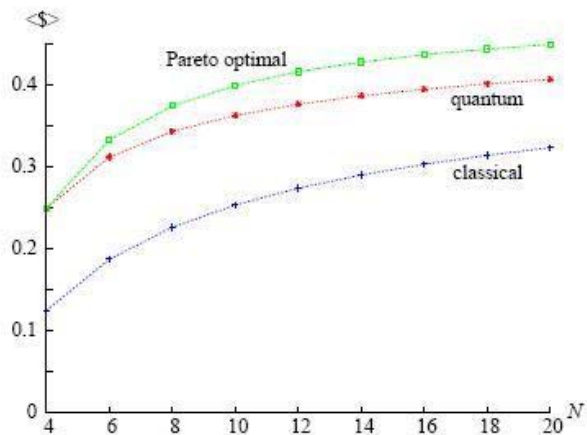

Fig. 5. The Nash equilibrium payoff as a function of the number of players ($N$) for even $N$ for the fully entangled quantum case (*) and the classical case (+). Compare with the Pareto optimal payoffs ( ). The curves slowly converge to $\langle \$ \rangle = \frac{1}{2}$ as $N \to \infty$.

## 4.1 The Many Games Interpretation of Quantum Worlds

So, after a rather roundabout journey from the bridge of the enterprise, we now have a number of quantum games with quantum Nash equilibria which are both uniquely distinguishable from the classical games and classical equilibria (Iqbal and Weigert, Cheon and Tsutsui, Flitney and Abbott, Flitney and Hollenberg, Landsberg) but we also have an interesting question with respect to quantum computing, which is what happens under conditions of decoherence.

Not unexpectedly, the general result of decoherence is to reduce the quantum Nash equilibrium to the classical Nash equilibrium, however, this does not happen in a uniform fashion. As Flitney and Hollenberg explain:

The addition of decoherence by dephasing (or measurement) to the four player quantum MG results in a gradual diminution of the NE payoff, ultimately to the classical value of 1/8 when the decoherence probability $p$ is maximized, as indicated in figure 6. However, the strategy given by Eq. (12) remains a NE for all $p < 1$. This is in contrast with the results of Johnson for the three player "El Farol bar problem" and ¨Ozdemir *et al.* for various two player games in the Eisert scheme, who showed that the quantum optimization did not survive above a certain noise threshold in the quantum games they considered. Bit, phase, and bit-phase flip errors result in a more rapid relaxation of the expected payoff to the classical value, as does depolarization, with similar behaviour for these error types for $p < 0.5$

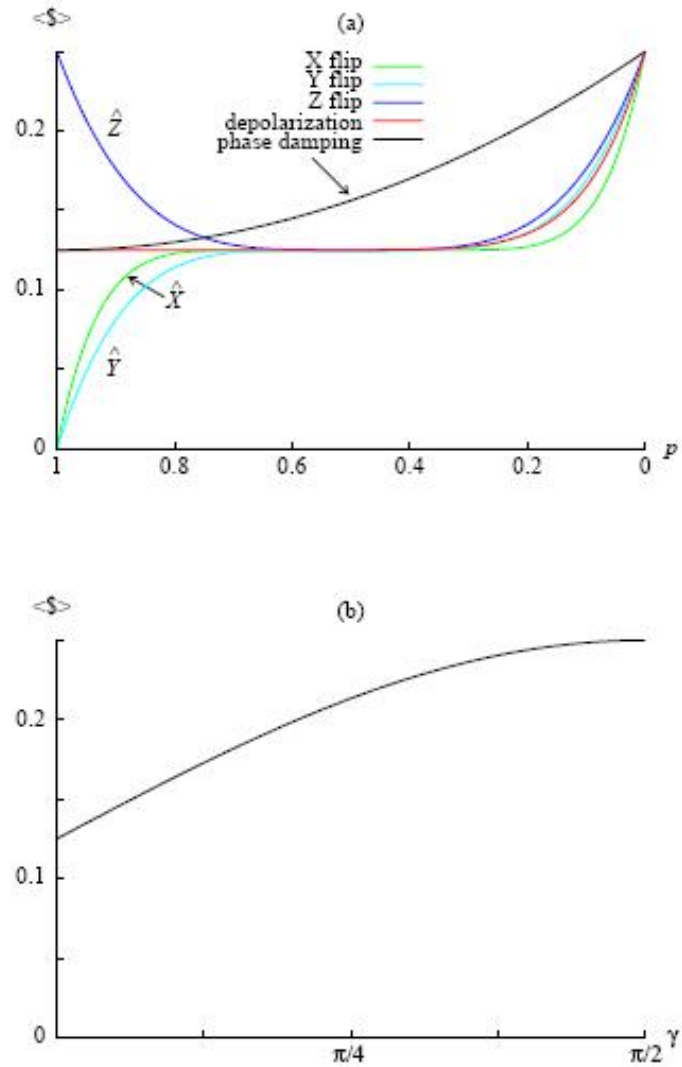

**Flitney and Hollenberg's Fig. 6** (a) The Nash equilibrium payoff in an $N = 4$ player quantum Minority game as a function of the decoherence probability $p$. The decoherence goes from the unperturbed quantum game at $p = 0$ (right) to maximum decoherence at $p = 1$ (left). The curves indicate decoherence by phase damping (black), depolarization (red), bit flip errors (green), phase flip errors (blue) and bit-phase flip errors (blue-green). Compare this with (b) the Nash equilibrium payoff for $N = 4$ as a function of the entangling parameter $\gamma$ [Eq. (11)].

## 5.1  Conclusion

Quantum computing does, indeed, give rise to new Nash equilibria, which belong to several different classes. Classical or apparently classical games assume new dimensions, generating a new strategy continuum, and new optima within and tangential to the strategy spaces as a function of quantum mechanics. A number of quantum games can also be mathematically distinguished from their classical counterparts and have Nash Equilibria different than those arising in the classical games. The introduction of decoherence, both as a theoretical measure, and perhaps, more importantly, as a performance measure of quantum information processing systems illustrates the ways in which quantum Nash equilibria are subject to conditions of "noise" and system performance limitations. The decay of higher optimality quantum Nash equilibria to classical equilibria is itself a complex and non-linear process following different dynamics for different species of errors. Finally, non-locality of the EPR type, and bearing an as yet incompletely understood relationship to Bell's Theorem offers a way in which quantum communication can be introduced into a variety of game theoretic settings, including both strategy and cryptography, in ways which profoundly modify attainable Nash equilibria. While the field has been slow to develop and most of the foundational research has come from a relatively small number of advances, the insights offered by these advances are profound and suggest that quantum computing will radically impact the fields of decision-making and communications in the near future.

# Appendix I: Huberman and Hogg's Coordinated Rock Paper Scissors Game[5]

Another example is a three-player version of the game of rock, paper, scissors, in which the two allied players must make the same choice to have any chance of winning. If the allies make different choices their payoffs are zero and the third player gets a payoff of 1. When the two allies make the same choice the payoff to the allies and the third player are given the payoff matrix of the usual two-player rock, paper scissors game, which is shown in Table 3. This game has the feature that no single choice is best, i.e., there is no pure strategy Nash equilibrium. Instead, the best strategy for rational players is to make the choices randomly and with equal probability, which gives it a mixed strategy Nash equilibrium with expected payoff of 1/3. For the full game without coordination the pair of allied players only has 1/3 chance of making the same choice, and another 1/3 to win against their opponent, leading to an expected payoff of 1/9. If they can be perfectly coordinated their payoff would be 1/3. In this example it is necessary to play random choices because any *a priori* commitment between the allied pair to a specific set of choices would no longer be a random strategy, and therefore discoverable by observation.

| Choice | Rock | Paper | Scissors |
|--------|------|-------|----------|
| Rock | 0, 0 | −1, 1 | 1, −1 |
| Paper | 1, −1 | 0, 0 | −1, 1 |
| Scissors | −1, 1 | 1, −1 | 0, 0 |

*Payoff structure for the Rock, Paper, Scissors for the pair of allied players against the third player. Each row and column corresponds to choices made by the pair (assuming they are the same) and their opponent respectively, and their corresponding payoffs. For example, the entry of the second column, second row corresponds to the allies both choosing paper and the third player choosing rock (p. 424.)*

---

[5] Taken from Huberman, B. and Hogg, T. (2004) "Quantum Solution of Coordination Problems", Quantum Information Processing, Vol. 2, No. 6, December, 2004

## Appendix II: The Minority Game (Agent Based Modeling) – J. Doyne Farmer[6]

The minority game represents the opposite end of the spectrum. Despite its simplicity, it displays some rich behavior. While the connection to markets is only metaphorical, its behavior hints at the problems with the traditional views of efficiency and equilibrium. The minority game was originally motivated by Brian Arthur's El Farol problem. El Farol is a bar in Santa Fe, near the original site of the Santa Fe Institute, which in the old days was a popular hangout for SFI denizens. In the El Farol problem, a fixed number of agents face the question of whether or not to attend the bar. If the bar is not crowded, as measured by a threshold on the total number of agents, an agent wins if he or she decides to attend. If the bar is too crowded, the agent wins by staying home. Agents make decisions based on the recent record of total attendance at the bar. This problem is like a market in that each agent tries to forecast the behavior of the aggregate and that no outcome makes everyone happy.

The minority game introduced by Damien Challet and Yi-Cheng Zhang is a more specific formulation of the El Farol problem. At each timestep, N agents choose between two possibilities (for example, A and B). A historical record is kept of the number of agents choosing A; because N is fixed, this automatically determines the number who chose B. The only information made public is the most popular choice. A given time step is labeled "0" if choice A is more popular and "1" if choice B is more popular. The agents' strategies are lookup tables whose inputs are based on the binary historical record for the previous m timesteps. Strategies can be constructed at random by simply assigning random outputs to each input (see Table A).

Each agent has s possible strategies, and at any given time plays the strategy that has been most successful up until that point in time. The ability to test multiple strategies and use the best strategy provides a simple learning mechanism. This learning is somewhat effective—for example, asymptotically A is chosen 50% of the time. But because there is no choice that satisfies everyone—indeed, no choice that satisfies the majority of the participants—there is a limit to what learning can achieve for the group as a whole.

Table A.

| Input | Output |
|-------|--------|
| 0 0   | 1      |
| 0 1   | 0      |
| 1 0   | 0      |
| 1 1   | 1      |

Example of a strategy for the minority game. The input is based on the attendance record for the m previous time-steps, 0 or 1, corresponding to which choice was most popular. In this case m = 2. The output of the strategy is its choice (0 or 1). Outputs are assigned t random.

When s > 1, the sequence of 0s and 1s corresponding to the attendance record is aperiodic. This is driven by switching between strategies. The set of active strategies continues to change even though the total pool of strategies is fixed. For a given number of agents, for small m the game is efficient, in that prediction is impossible, but when m is large, this is no longer the case. In the limit N→ ∞, as m increases there is a sharp transition between the efficient and the inefficient regime.

The standard deviation of the historical attendance record, σ, provides an interesting measure of the average utility. Assume that each agent satisfies his or her utility function by making the minority choice. The average utility is highest when the two choices are almost equally popular. For example, with 101 agents the maximum utility is achieved if 50 agents make one choice and 51 the other. However, it is impossible to achieve this state consistently. There are fluctuations around the optimal attendance level, lowering the average utility. As m increases, σ exhibits interesting behavior, starting out at a maximum, decreasing to a minimum, and then rising to obtain an asymptotic value in the limit as m→ ∞. The minimum occurs at the transition between the efficient and inefficient regimes. The distinction between the efficient and inefficient regimes arises from the change in the size of the pool of strategies present in the population, relative to the total number of possible strategies. The size of the pool of strategies is sN.

---

[6] Excerpted from "J. Doyne Farmer, "Physicists Attempt to Scale the Ivory Towers of Finance", Computing in Science and Engineering, November-December, 1999.

The number of possible strategies is $_2 2^m$, which grows extremely rapidly with m. For example, for m = 2 there are 16 possible strategies, for m = 5 there are roughly 4 billion, and for m =10 there are more than 10$^{300}$ —far exceeding the number of elementary particles in the universe. In contrast, with s = 2 and N = 100, there are only 200 strategies actually present in the pool. For low m, when the space of strategies is well-covered, the conditional probability for a given transition is the same for all histories—there are no patterns of length m. But when m is larger, so that the strategies are only sparsely filling the space of possibilities, patterns remain. We can interpret this as meaning that the market is efficient for small m and inefficient for large m. The El Farol problem and minority game is a simple game with no solution that can satisfy everyone. This is analogous to a market where not everyone profits on any given trade. *Studies of the minority game suggest that the long-term behavior is aperiodic: the aggregate behavior continues to fluctuate. In contrast to the standard view in economics, such fluctuations occur even in the absence of any new external information.*

## Appendix III: Flitney and Hollenberg's Decoherence Measurement Scheme[7]

The physical implementation of a quantum system determines when the decoherence operators should be inserted in the formalism. For example, in a solid state implementation, errors, including qubit memory errors, need to be considered after each time step, while in an optical implementation memory errors only arise from infrequent photon loss, but errors need to be associated with each quantum gate. In addition, there may be errors occurring in the final measurement process. In this paper we shall describe a quantum game in the Eisert scheme with decoherence in the following manner (Equation 7):

$$
\begin{array}{lll}
\rho_0 = & |\psi_0\rangle\langle\psi_0| & \text{(initial state)} \\
\rho_1 = & \hat{J}\rho_0\hat{J}^\dagger & \text{(entanglement)} \\
\rho_2 = & D(\rho_1, p) & \text{(partial decoherence)} \\
\rho_3 = & (\otimes_{k=1}^N \hat{M}_k)\,\rho_2\,(\otimes_{k=1}^N \hat{M}_k)^\dagger & \text{(players' moves)} \\
\rho_4 = & D(\rho_3, p') & \text{(partial decoherence)} \\
\rho_5 = & \hat{J}^\dagger\rho_4\hat{J} & \text{(preparation for measurement)}
\end{array}
$$

to produce the final state $p_f \equiv p_5$ upon which a measurement is taken. That is, errors are considered after the initial entanglement and after the players' moves. In all subsequent calculations we set $p' = p$. An additional error possibility could be included after the $\hat{J}^\dagger$ gate but this gate is not relevant in the quantum Minority game since it only mixes states with the same player(s) winning. Hence the gate and any associated decoherence will be omitted for the remainder of the paper. The $\hat{J}$ gate can be implemented by a (generalized) Hadamard gate followed by a sequence of CNOT gates, as indicated in figure 1.

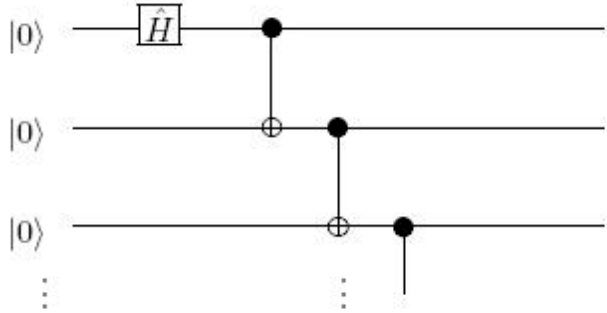

Fig. 1. A possible gate sequence to implement the entangling $\hat{J}$ gate.

When the number of qubits is large the possibility of errors occurring within the J gate needs to be considered but is not done so here. The function **D(p; ρ)** is a completely positive map that applies some form of decoherence to the state **ρ** controlled by the probability **p**. For example, for bit flip errors

$$
D(\rho, p) = (\sqrt{p}\,\hat{\sigma}_x + \sqrt{1-p}\,\hat{I})^{\otimes N}\,\rho\,(\sqrt{p}\,\hat{\sigma}_x + \sqrt{1-p}\,\hat{I})^{\otimes N}
$$

The scheme of Equation 7 is shown in Figure 2.

---

*Multiplayer quantum Minority game with decoherence*

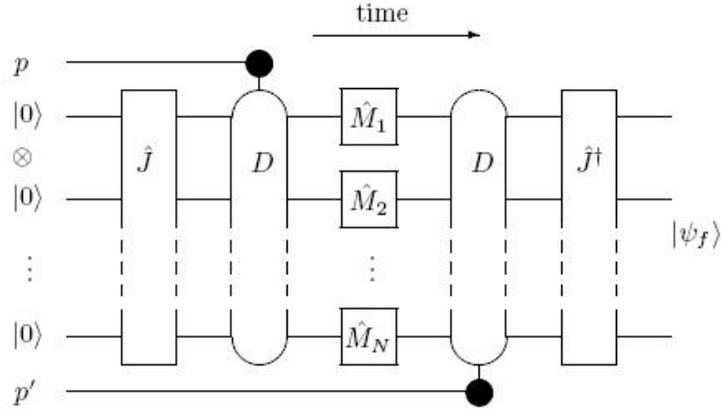

Fig. 2. The flow of information in an $N$-person quantum game with decoherence, where $M_k$ is the move of the $k$th player and $\hat{J}$ ($\hat{J}^\dagger$) is an entangling (dis-entangling) gate. The central horizontal lines are the players' qubits and the top and bottom lines are classical random bits with a probability $p$ or $p'$, respectively, of being 1. Here, $D$ is some form of decoherence controlled by the classical bits.

The expectation value of the payoff to the k$^{\text{th}}$ player is:

$$\langle \$^k \rangle = \sum_{\xi} \hat{\mathcal{P}}_\xi \, \rho_f \, \hat{\mathcal{P}}_\xi^\dagger \, \$_\xi^k, \tag{9}$$

where $\hat{\mathcal{P}}_\xi = |\xi\rangle\langle\xi|$ is the projector onto the computational state $|\xi\rangle$, $\$_\xi^k$ is the payoff to the $k$th player when the final state is $|\xi\rangle$, and the summation is taken over $\xi = j_1 j_2 \ldots j_N$, $j_i \in \{0,1\}$.